# Bonded straight and helical flagellar filaments form ultra-low-density glasses

Sevim Yardimci[1,2,*], Thomas Gibaud[1,3*], Walter Schwenger[1], Matthew R. Sartucci[4], Peter D. Olmsted[4], Jeffrey S. Urbach[4], Zvonimir Dogic[5,6,1]

[1]The Martin Fisher School of Physics, Brandeis University, Waltham, Massachusetts 02454, USA
[2]The Francis Crick Institute, NW1 1AT London, UK.
[3]Univ Lyon, Ens de Lyon, Univ Claude Bernard, CNRS, Laboratoire de Physique, F-69342 Lyon, France
[4]Department of Physics and Institute for Soft Matter Synthesis and Metrology, Georgetown University, Washington, DC 20057, USA.
[5]Department of Physics, University of California at Santa Barbara, Santa Barbara, CA, 93106, USA.
[6]Biomolecular Science and Engineering, University of California at Santa Barbara, Santa Barbara, CA, 93106, USA.

**Abstract:** We study how the three-dimensional shape of rigid filaments determines the microscopic dynamics and macroscopic rheology of entangled semi-dilute Brownian suspensions. To control the filament shape we use bacterial flagella, which are micron-long helices assembled from flagellin monomers. We compare the dynamics of straight rods, helical filaments, and shape diblock copolymers composed of seamlessly joined straight and helical segments. Caged by their neighbors, straight rods preferentially diffuse along their long axis, but exhibit significantly suppressed rotational diffusion. Entangled helical filaments escape their confining tube by corkscrewing through the dense obstacles created by other filaments. By comparison, the adjoining segments of the rod-helix shape-diblocks suppress both the translation and the corkscrewing dynamics, so that shape-diblocks become permanently jammed at exceedingly low densities. We also measure the rheological properties of semi-dilute suspensions and relate their mechanical properties to the microscopic dynamics of constituent filaments. In particular, rheology shows that an entangled suspension of shape rod-helix copolymers forms a low-density glass whose elastic modulus can be estimated by accounting for how shear deformations reduce the entropic degrees of freedom of constrained filaments. Our results demonstrate that the three-dimensional shape of rigid filaments can be used to design rheological properties of semi-dilute fibrous suspensions.

**Introduction:** Macroscopic properties of diverse soft materials are largely determined by the shape of their constituents. For example, the shape of granular particles dramatically influences the properties of the jamming transition [1-6]. In microscopic systems where thermal noise becomes relevant, changing the particle shape from spheres to polyhedra to ellipsoids determines the packing, phase behavior, and rheology of dense suspensions [7-11]. Filamentous molecules are another class of soft materials whose macroscopic properties are critically influenced by their shape; they are usually classified into three categories according to the backbone rigidity: flexible, semi-flexible, and rigid polymers [12,13]. The shape, connectivity, and topology of flexible chains dramatically influence both the microscopic dynamics and macroscopic rheology of polymeric suspensions. Branched, ring and star-shaped flexible polymers have properties that are very different from their linear analogs [14-16]. Flexible polymers seamlessly change between many accessible conformations. In comparison, filaments with large backbone rigidity are permanently frozen in one particular conformation. Studies of rigid filaments have largely focused on one type of molecule, straight rods with no intrinsic curvature [17-20]. However, one can envision other types of rigid filaments, such as helices, crosses, or L-shaped molecules which have permanently frozen non-zero curvature that determines their 3D shape. Theoretical efforts have started to explore how the shape of such 3D rigid filaments affects the properties of their suspensions [21-25]. However, progress in this area has been limited by the lack of an experimental system in which one can tune the shape of rigid filaments from previously studied straight rods to more complex meandering shapes.

We study microscopic dynamics and macroscopic rheology of entangled semi-dilute suspensions composed of flagellar filaments. Bacterial flagella can assume several distinct helical and straight polymorphic shapes [26]. Furthermore, these shapes can be sequentially joined to each other to form composite multi-block shape copolymers [27]. Using this system we studied suspensions of straight and helical filaments. We have also sequentially polymerized straight and helical segments to create rod-helix shape diblock-copolymers. Increasing the shape complexity increasingly constrains the dynamics of filaments in semi-dilute isotropic suspensions. We related such microscopic dynamics to the rheological properties of the bulk suspensions. Rod-helix diblock copolymers become permanently jammed at exceedingly low volume fractions, leading to a new route for assembling low-density soft solids.

**Prokaryotic flagella as a model for shape-tunable rigid filaments:** Flagella are long helical filaments that are attached to the surface of the bacterium by a rotary molecular motor [28,29]. Flagellar filaments are hollow cylinders that are assembled from 11 laterally associating linear protofilaments [30,31]. In turn, protofilaments are assembled from flagellins, a monomeric protein that exists in two distinct conformational states [32]. Depending on suspension conditions and mutations in amino-acid flagellin sequence, a variable fraction of adjacent protofilaments is in two conformational states, generating incommensurate lateral binding sites [33,34]. The resulting frustration is resolved by the formation of 12 distinct states, ten of which are helical and two straight.

Flagellar filaments have a ~20 nm diameter. The average contour length $l_c$ of the filaments used in our study was several microns. Flagella have a persistence length of $l_p$ ~1 mm [35,36]. Since $l_p \gg l_c$, we consider flagella to be rigid filaments. We studied suspensions of flagella with three distinct shapes: First, we prepared straight filaments isolated from *Salmonella Typhimurium* strain SJW1660 (**Fig. 1a**). Second, we purified flagella from *Salmonella Typhimurium* strain SJW1103 (**Fig. 1b**). These filaments assumed a helical superstructure with $a_\lambda$~0.75 μm helical diameter and $\lambda$ ~2.4 μm helical pitch (**Fig. 1c**) [37]. The straight and helical filaments differ by a single point mutation in the amino-acid sequence of the flagellin monomer [38]. Third, we prepared rod-helix shape diblock-copolymers comprising adjoining straight and helical segments (**Fig. 1d**). To assemble such filaments we first polymerized a segment of straight filaments isolated from SJW1660 strain. Subsequently, we used these segments as seeds to induce the polymerization of flagellin monomers that form wild-type helices and were isolated from SJW1103 strain. The local curvature lacks any discontinuities as the straight segment smoothly transitioned into a helical one. The angle $\alpha$ between the axis of the straight filament and the long axis of the curly filament was measured to be: $\alpha = 128 \pm 9$, which is close to the value of $\alpha = 136$, which is predicted from the parameters of the flagellar helix.

**Filament shape determines filament dynamics:** To investigate how the filament shape determines microscopic dynamics we prepared semi-dilute isotropic suspensions of straight, helical, and rod-helix shape diblock filaments at 5 mg/ml. These samples were doped with a low volume fraction of labeled filaments of the same shape, and imaged with fluorescence microscopy (**Fig. 2**). Overlaying sequential images of diffusing fluorescent filaments revealed qualitative differences in the dynamics of three shapes.

Similar to flexible polymers, adjacent elongated particles confine each straight filament within an effective tube [17,39,40]. On short-time scales this confinement suppressed both the filament's rotation and its translation in the directions perpendicular to its long axis (**Fig. 2a**). In contrast, the motion of the rigid rod along its long axis remains unobstructed. On longer time scales the filament escapes its confining tube by translating along its long axis and rotating around its center of mass. Such reptation-like dynamics are revealed by plotting a sequence of filament configurations on top of each other, where a temporal evolution is marked by a color gradient (**Fig. 2a**). Similar dynamics have been observed in other rod-like systems including nanotubes in a dense gel, needles in 2D obstacles and computer simulations [41-43].

Helical filaments exhibit different dynamics. Because of the limited resolution along the optical axis, helical filaments appeared as sinuous curves whose adjacent peaks and valleys are separated by half the helical pitch (**Fig. 2b**). Pure translation of the helix's center of mass along the filament's long axis translates the locations of the peaks and valleys. Pure rotation about the long helical axis fixes the center of mass while translating the positions of the peaks and valleys. Superimposing a temporal sequence of helix positions reveals that the peaks remained fixed through the observation time, yet the helix's center of mass diffuses back and forth (**Fig. 2b**). This observation suggests that a helix is also confined to an effective tube. However, such tube is not a simple cylinder but rather a helix whose shape is determined by the filament geometry. As a confined helix undergoes a pure translation along its long axis, it quickly encounters obstacles due to other filaments; thus such motion is suppressed. Similarly, pure helical rotation is also suppressed. The only mode by which the helix can escape its local cage is the corkscrewing motion along its long axis, where a translation by one-pitch length corresponds to a $2\pi$ rotation. Because such rotations result in the net translation of the center of mass of the helix, the corkscrewing dynamics allow helices to escape their confining tube. The dynamics of helical flagella in entangled isotropic solutions are similar to that observed in twist-bend liquid crystalline phases [44].

To quantify the microscopic dynamics we tracked the motion of fluorescently labeled helices diffusing within a semi-dilute isotropic suspension of unlabeled filaments. We focused on image sequences where helices remained in the image plane. We defined the filament reference frame, where the coordinates $(r_\parallel, r_\perp)$ are aligned along the long axis of the helix in the initial image (**Fig. 3a**). For times smaller than the reptation time one can describe the position of the helix by its center of mass $(\Delta r_\parallel, \Delta r_\perp)$. Diffusion along the filament's long axis was linear for measured time scales, with an effective diffusion constant of $D_\parallel = 0.08 \pm 0.01$ μm²/s where we assumed: $\langle \Delta r_\parallel^2 \rangle = 2D_\parallel \Delta t$ (**Fig. 3b**). To characterize coupling between translation and rotation we measured the helix's phase angle $\phi(t)$ as a function of time. The phase angle can be translated into distance according to $r_\phi(t) = \lambda \phi(t)/2\pi$, where $\lambda$ is the helix pitch (**Fig. 3a**). The two displacements $\Delta r_\parallel(t)$, and $r_\phi(t)$ track each other (**Fig. 3c**), as expected for pure corkscrewing dynamics; i.e. the center of mass does not proceed by rigid diffusion, even at the smallest timesteps observed.

The rod-helix shape-diblocks exhibited microscopic dynamics distinct from both straight and helical filaments. Overlaying shape-diblock contours taken over 200 seconds revealed minimal displacement, suggesting strong geometrical confinement (**Fig. 2c**). This observation can be understood as follows: The straight segment of the diblock copolymer can escape its confining tube only by translating along its long-axis, while the helical segment moves by the corkscrewing motion. These two dynamical modes are fundamentally incompatible due to the rigid nature of the bond between the two segments. Any corkscrewing motion of the helical segment would cause the straight segment to sweep out a wide cone that is microns in diameter; this is impossible because of the neighbouring filaments. Similarly, a simple translation of the straight segment along its long axis is incompatible with the corkscrewing motion of the helical segment. Thus, rod-helix shape-diblocks are permanently trapped in their confining tube.

We quantified the dynamics of shape-diblocks at different concentrations (**Fig. 4**). Measured MSD curves associated with both the translation and orientational degrees of freedom remained sub-diffusive at all measured timescales for filament concentrations around 2 mg/ml (**Fig. 4a**). Increasing the concentration of shape-diblocks from 2 to 5 mg/ml decreased the magnitude of the MSD displacements for all measured times. We used the measured MSD curves to estimate the effective size of the tube that cages the diblock copolymers. We defined the effective diameter $\sigma$ of the cage as the square root of a mean square displacement at a particular time lag ($\Delta t = 10$ s). With decreasing filaments concentration the effective cage size increased rapidly (**Fig. 4c, d**). Below critical filaments concentration of ~1.5 mg/ml the cage size diverges which suggests that the sample fluidizes.

**Rheology of helical and rod-helix diblock filaments:** The microscopic dynamics of the constituent filaments is intimately related to the macroscopic rheological properties of entangled semi-dilute suspensions. To explore this relationship we characterized the rheology of flagellar filaments. Imposing an oscillatory strain of amplitude $\gamma$ at frequency, $f$, we measured the storage $G'(f)$ and the loss $G''(f)$ moduli. For a given sample we varied the frequency while keeping the amplitude constant and small enough to remain in the linear response regime. The modulus of the suspension of straight filaments at 5 mg/ml was below the sensitivity limit of our rheometer, while above 5 mg/ml straight filaments formed a nematic liquid crystal [44], which alters the suspension's rheological properties.

In contrast to straight rods, the viscoelasticity of helical filaments was significantly larger and thus easily measured with the rheometer. Below 6.0 mg/ml concentrations the loss modulus $G''(\omega)$ was either comparable to or larger than the storage modulus $G''(\omega)$ for all frequencies. For 6.0 mg/ml and higher concentrations, the moduli exhibited three distinct regimes separated by two crossover frequencies, $f_{LF}$ and $f_{HF}$ (**Fig. 5a**). Fluid-like behavior ($G'' > G'$) was observed at frequencies below $f_{LF}$. In the high-frequency regime, above $f_{HF}$, the loss modulus also dominated the storage modulus. At intermediate frequencies the elastic modulus dominated over the viscous one ($G' > G''$). These distinct regimes are typical of viscoelastic fluids such as polymeric suspensions or wormlike micelles [12]. The high-frequency regime probes the dynamics of single filaments, while relaxation at low frequencies occurs due to filaments escaping their cages through the corkscrew-reptation mechanism.

With increasing flagella concentration the magnitude of the elastic modulus increased significantly (**Fig. 5b**). For example, the elastic modulus at the lower crossover frequency $G(f_{LF})$ increased from 0.1 to 2 Pa as the concentration increased from 6.6 to 16 mg/ml. The lower crossover frequency systematically shifted towards lower values with increasing concentration, from $f_{LF} = 0.2$ Hz at 6.0 mg/ml to $f_{LF} \sim 0.04$ Hz at 16 mg/ml (**Fig. 5c**). This suggests that the time for the filament to escape its confining tube due to corkscrew-reptation is $\tau_{CR} \sim 5$-$20$ s, which is comparable to the time scales observed from microscopy images.

Next, we examined the viscoelastic properties of rod-helix diblocks (**Fig. 5d**). For dilute samples at or below 2 mg/ml the viscous modulus dominates the elastic one for all frequencies. At higher concentrations, an elastic regime emerges at low frequencies. Similar to helical filaments, the loss modulus dominates at high frequencies ($G'' > G'$), with a cross-over to an elastic regime below $f_{HF}$. However, in contrast to helical filaments, we did not observe a lower cross-over frequency to a fluid regime; instead, the elastic modulus dominated for all measured frequencies below $f_{HF}$. At the highest measured concentration of 4 mg/ml we observed a nearly frequency-independent elastic plateau of $G' \sim 0.7$ Pa, more reminiscent of a cross-linked rubber than a viscoelastic polymer liquid. This rheological behavior is consistent with the observed single filament dynamics showing that shape-diblocks are permanently caged. Upon an imposed strain deformation the filaments cannot relax into a lower energy state. Notably, the elastic modulus for shape-diblocks was significantly larger than the modulus of helical filaments at comparable concentrations.

Having measured both the microscopic dynamics of shape diblock and their rheological properties make it possible to relate the two. Scaling arguments suggest that the elastic modulus scales with the effective volume that is accessible to the caged filaments. We find that the inverse of the cage size, estimated from the mean-square displacement of fluorescently labeled filaments, scales linearly with the independently measured elastic modulus (**Fig. 6**). Importantly, this relationship holds for several elastic moduli measured at several frequencies and estimating the cage size from the corresponding times in the MSD curves.

**Origin of elasticity of shape block copolymers:** The reptation dynamics of individual filaments qualitatively explain the shape of the low-frequency rheological response of both helical and shape-diblock suspensions. Quantitative understanding of the low-frequency elastic modulus of the shape-diblocks is more challenging. The elastic modulus for perfectly rigid sterically interacting filaments, arising solely from the orientational degrees of freedom, is of the order:

$$G'_{orient} \sim c_n k_B T \tag{1}$$

where $c_n = cL/\rho_L$ is the filament number density [12,45]. The mass per unit length for the flagella is $\rho_L \approx 1.8 \times 10^{-13}$ mg/μm, and the average filament length is $L \approx 6.1 \pm 1.2$ μm, [46]. For concentration of c = 4 mg/ml, the filament number density is $c_n \sim 4.4 \times 10^{12}$ filaments/ml. It follows that G'$_{orient}$~0.02 Pa, which is substantially lower than the measured plateau modulus for the suspension of shape-diblocks.

Another source of elasticity was proposed in a recent analysis of wire frame glasses, which are dense suspensions of particles composed of rigid straight segments connected by a joint of a fixed angle [23,24]. Just as with the diblock particles studied here, the rigid joint suppresses reptation and leads to the formation of a glass-like solid with increasing concentration [25]. At high concentrations, the low-frequency elastic response for the L-shaped wire-frame particles is dominated by entropic effects due to the dramatic reduction in the cage size that occurs when an equilibrium suspension is strained. Geometric arguments show that a typical particle becomes completely constrained at a strain of $\gamma_c \sim (c_n L^3)^{-1}$, at which point its entropy becomes zero. Further strain requires particle deformations, with a corresponding contribution to the free energy determined by the filament bending energy.

The approximate strain-dependent free energy and the associated elastic modulus of a wire frame glass arising from the above-described entropic effect are given by the following expressions:

$$F/k_B T \sim \frac{c_n}{\gamma_c^2} \gamma^2 (1 - (\gamma/\gamma_c)^2) \tag{2}$$

$$G = \frac{1}{\gamma}\frac{\partial F}{\partial \gamma} \sim \frac{2\, c_n\, k_B T}{{\gamma_c}^2}(1 - 2(\gamma/\gamma_c)^2) \tag{3}$$

These expressions highlight the important role of the critical strain $\gamma_c$, in the properties of completely entangled rigid filaments. The first term in Eq. 3 determines the linear modulus, which accounts for the rate at which the entropy decreases as the suspension is strained. This term shows that the elastic modulus of wire-frame glasses is enhanced by the factor of $\frac{1}{\gamma_c^2}$, when compared to an equivalent suspension of rigid rods (Eq. 1). The second term in Eq. 3 shows that wire-frame solids exhibit non-linear strain softening, and the rate of strain softening is controlled by the critical strain $\gamma_c$. The expression for elastic modulus (Eq. 3) does not include the contribution from filament bending energy, which produces nonlinear strain stiffening [24].

The measured non-linear strain-dependent elasticity suggests that the model derived for wireframes can explain the origin of the measured elasticity for the diblock particles (**Fig. 6**). Equation 3 describes the initial strain-softening of the elastic modulus. We estimate $\gamma_c$ by identifying the strain $\gamma_1$ at which the modulus drops by 10% below its value in the linear regime, which, from Equation 3, implies that $\gamma_c = \sqrt{20}\gamma_1$. The linear modulus is determined at the strain amplitude of $\gamma = 0.01$ by interpolating the strain-dependent elastic modulus (Fig. 6). We find $\gamma_c \sim 0.24$ and $0.46$ for samples at 4 mg/ml and 3 mg/ml, respectively. For lower concentrations, the modulus was too small to accurately determine strain softening. Thus, the modulus for 4 mg/ml suspension will increase by a factor of ~17 relative that that of straight rods at similar concentrations, and a factor of ~5 for 3 mg/ml samples. Both the modulus magnitude and the trend with concentration are reasonably well described by the wireframe model. In principle, the critical strain could also be estimated from the relationship $\gamma_c \sim (c_n L^3)^{-1}$. However, this relationship significantly underestimates the magnitude of the critical strain. The critical strain is highly sensitive to filament length and the theoretical model is derived for a suspension of uniform-length filaments. In comparison, flagellar filaments have significant polydispersity that could strongly influence how the critical strain depends on the filament's geometry.

The absence of strain stiffening in the measured amplitude sweeps suggests that the bending energy of the shape-diblocks is small (**Fig. 6**). Specifically, the wire glass model predicts that the enthalpic contribution to the nonlinear rheology, determined by the energy $E_{bend} = \frac{K\alpha^2}{2}$ to bend by an angle $\alpha$, produces strain stiffening and dominates the entropic strain softening when $K \geq G/c_n$. A non-zero contribution from $K$ would affect the nonlinear term in equation 3, but is not considered here. The value of $K$ depends on the bending modulus of the filament, which has been measured [34], but the relationship for the complex geometry of the diblocks is not straightforward and likely depends on other filament properties that have not been directly measured.

**Discussion:** Our work demonstrates that the shape of three-dimensional rigid filaments dramatically affects both the dynamics and the rheology of bulk suspensions. Filaments with straight, helical, and rod-helix shapes exhibit increasingly constrained microscopic dynamics, which is also reflected in the measured rheological properties. With increasing concentration, the rod-helix shape-diblock filaments became increasingly constrained at all measured timescales. Consequently, they exhibit a solid glass-like state even in the absence of chemical crosslinkers.

The formalism of wire-frame particles provides a potential framework for understanding the mechanics of fully constrained suspensions of bent or curved filaments that have elastic plateaus at very low frequencies. Unlike shape-diblocks, helical filaments are not fully constrained and thus their elastic moduli vanish for very low frequencies. Nevertheless, at intermediate frequencies, helical filaments exhibit a larger elastic plateau than rod-like particles of comparable length and the same concentrations. Thus, our results demonstrate the need to develop theoretical models that can bridge the limits of previously studied rigid rods and the emerging formalism of fully constrained systems.

Besides flagella, other emerging technologies make it possible to engineer the shape of rigid filaments. For instance, recent advances in the DNA origami field enable the design of particles of almost any shape and their purification in large enough quantities to perform bulk rheological experiments [47,48]. Taken together, these advances form the basis for rationally engineering the mechanics of soft materials by controlling the particle's three-dimensional shape. They also demonstrated the need to develop a theoretical formalism to rigorously connect the 3D filament shape to the microscopic dynamics and the associated macroscopic rheological properties.

Our results demonstrate the promise of flagellar filaments as a model system to explore the suspension of rigid filaments. Flagellar filaments allow for visualizing the microscopic dynamics at a single filament level, which can then be correlated with macroscopic rheology. Previous work has shown helical filaments at high densities form a liquid crystalline phase [44]. Besides exhibiting well-defined helical and straight shapes, flagellar filaments exhibit unique polymorphic transitions, where the entire structure can switch between distinct forms in response to diverse external stimuli such as chemical, temperature, or external stress [37,49,50]. When coupled with the shape-dependent rheology, controlling shape transitions provides a unique experimental platform wherein Angstrom-sized stimuli-induced conformational changes at a single monomer level [32,49] cascade over several hierarchical scales to change the filament shape and thus control the shape and macroscale mechanical properties.

**Acknowledgments:** S.Y, W. S., and Z.D acknowledge the support of the National Science Foundation through grant: NSF-DMR-1905384. J.S.U and M.R.S acknowledge the support of the National Institute for Standards and Technology through grant NIST-70NANB14H214. P.D.O. acknowledges the support of the Ives Foundation. We also acknowledge the support of Brandeis MRSEC biosynthesis and optical microscopy facility supported by NSF-MRSEC- 2011846, and the US–Israel Binational Science Foundation (BSF) through grant number 2016311.

**Materials and methods**

***Purification of flagella:*** Purification of flagella was carried out by previously described methods [51,52]. Helical flagella were purified from strain SJW1103, and straight flagella were purified from strain SJW1660. An isolated colony from an agar LB plate was grown in a 3 ml starter culture containing LB media. The initial culture was mixed with 200 ml LB media and incubated on a shaker at 37 °C until the optical density of the culture reached 0.6. This culture was then placed into 1 liter LB media to prepare large-scale cultures that were grown at 37°C until the optical density reached 0.6. Cells were spun at 10,000 g for 10 minutes to sediment the bacteria into a dense pellet. Subsequently, bacteria were collected in a falcon tube by diluting each pellet with only 1 ml, 10 mM phosphate buffer (pH 7.0). The dense bacteria suspension was vortexed for a few minutes. The resulting shear forces effectively separated bacterial bodies from their flagella.

Flagellar filaments were isolated from cell bodies by differential centrifugation. A concentrated bacteria sample was diluted 500-fold with 10 mM phosphate buffer (pH 7.0), and centrifuged at 12,000 g for 15 minutes at 4 °C to remove bacterial bodies. The presence of filaments in the supernatant was confirmed by visualization with dark-field microscopy. In the subsequent step, the supernatant was centrifuged at 100,000 g for 45 minutes at 4 °C to pellet the flagella. The pellet was re-suspended in 10 mM phosphate buffer. To further purify the filaments and separate them from other debris, filaments were bundled using PEG as a depletion agent. Specifically, 1 volume of PEG solution (20% w/w PEG MW 8000, 2.5 M NaCl) was mixed with 5 volumes of the suspension to introduce an osmotic pressure. Flagellar bundles were sedimented at 27,000 g for 10 minutes at 4 °C. Finally, flagella were re-suspended in 10 mM phosphate buffer. Flagella concentration was determined by measuring the sample optical density. To reduce the scattering flagellar filaments were de-polymerized by placing the sample in a heat bath at 65 °C for 10 minutes. Depolymerized samples showed a clear peak at 280 nm, and the protein concentration was estimated from an extinction coefficient $E$ (1%, 280nm) = 3.6 [53].

**Reconstitution of block copolymer filaments:** To prepare block copolymers we re-polymerized flagellar filaments from monomeric flagellin proteins [51,52,54-57]. The barrier for filament nucleation from flagellin monomer suspension is high; one never observes spontaneous nucleation under normal salt conditions. Instead, successful polymerization requires short seeds in the flagellin suspension. The polymerization of flagella is a delicate procedure as various contaminants from bacterial growth including capping proteins inhibit re-polymerization. Therefore, before repolymerization, the flagellin was purified by using ion exchange chromatography.

We polymerized flagellin monomers from short filaments, called 'seeds'. Tip-sonicator (Branson Sonifier) was used to break the filaments into short seeds, (2 minutes in total with 10 seconds pulses and 5 seconds breaks on ice). While 10% of the seeds were kept for nucleation, the rest of the fragments were depolymerized in a heat bath at 65°C for 10 minutes [51]. De-polymerized monomers were centrifuged at 346,000 g for 30 mins to clarify the solution. The supernatant containing flagellin was dialyzed overnight against 5 mM phosphate buffer (pH 7.0), using a 7,000 MW dialysis membrane. Anion exchange chromatography using Mono Q 5/50 GL was used with salt gradient for flagellin purification. The flagellin in 5 mM phosphate without salt was loaded on the chromatography column. The sample was eluted over a gradient of ionic strength up to 10 mM phosphate buffer (pH 7.0) with 0.5 M NaCl. The majority of the flagellin was eluted in fractions between 30 to 90 mM NaCl. Purified flagellin was analysed with gel electrophoresis that showed a clear band between 66,200 and 45,000 Da.

Reconstitution of block copolymer filaments was initiated by mixing 1 volume of 5 mg/ml seeds and 10 volumes of 5 mg/ml monomers from SJW1103 in 0.5 M NaCl, 10 mM phosphate buffer at neutral pH at room temperature. Incubation of this mixture for 5 hours led to the formation of helical filaments. An equal volume of 5 mg/ml monomers from SJW1660 was then added such that filaments continue to polymerize as straight filaments. After 5 hours of further incubation, filaments were centrifuged at 100,000 g for 45 minutes at 4 °C, the supernatant containing free monomers was removed, and filaments were resuspended in 10 mM phosphate buffer (pH 7.0).

*Fluorescence Labelling of Flagella and Imaging:* Flagellar filament has exposed amino groups on the exterior and can be labeled with an amine-reactive dye. 5-10 mg/ml flagellar suspension in 10 mM phosphate buffer (pH 7.0) was diluted 10-fold with sodium bicarbonate buffer at pH 8.8. The flagellar filament was labeled by mixing 20 μl of 1 mg/ml amine reactive rhodamine (5-(and-6)-Carboxy-tetramethylrhodamine-succinimidyl ester, Fisher Scientific) with 0.5 ml of 0.5-1 mg/ml flagellar suspension. Incubating the mixture at room temperature for 1 hour with continuous stirring was sufficient to reach maximum labeling. Free dye was separated from labeled filaments by dialysis or centrifugation. Rhodamine-labeled flagella were imaged with epi-fluorescence microscopy using 543 nm wavelength light and images were acquired on a Nikon inverted microscope equipped with an oil-type objective (Nikon Ti-2000, 100x, NA=1.49) with a CCD camera (Andor Neo). Data were acquired at three different rates. For long time scales (more than 1 minute), 50 ms exposure time was used with 1-second intervals. For shorter time scales, data were acquired at 10 ms or 50 ms exposure without delay.

*Data Analysis:* ImageJ was used to threshold and skeletonize microscopy images [8]. Subsequently, a custom IDL code was used to determine the coordinates of the contour length of the filaments with sub-pixel precision. Lastly, we used MATLAB code to analyze the dynamics of the flagella based on the coordinates of its contour length. To measure the mean-square displacement (MSD), we tracked five to ten filaments and averaged all the results over these runs. Although the flagella samples are quite polydisperse we only tracked filaments that were close to the average length of the filaments in the solution for 50-100 seconds. To quantify filament dynamics, we determined the angle $\theta$ between the x-axis of the laboratory and the $r_{//}$-axis defined by the long axis of the filament. The center of mass of the filament ($r_{//}$, $r\perp$) was then tracked in the referential ($r_{//}r\perp$) defined by the orientation of the filament. $\phi$ is the phase angle that the filament makes around its axis. We determined the phase angle by fitting a sinusoidal wave to the flagella. Because the image acquisition rate is high, the phase shift $\delta\phi=\phi(\ +1)-\phi(t)$ between two successive images is small ($\delta\phi<<180°$). We, therefore, select a fitting value $\phi$ that $\delta\phi$ remains small and that $\phi$ is a continuous and smooth function of time.

**Rheological Measurements:** The mechanical properties of flagellar suspensions were measured with a constant strain rheometer (ARES G2, TA instruments); with a 50 mm diameter, 0.0195 RAD stainless steel cone-plate geometry, and a 51.8 μm truncation gap. All measurements were performed at 25°C. The average total contour length of the shape diblock filaments was 6.1 $\pm$ 1.2 $\mu$m, while the average length of the straight segment was 2.7 $\pm$ 1.1 $\mu$m. The average length of the helical filaments was 6.1 $\pm$ 1.2 $\mu$m. To prevent sample loss due to evaporation the sides of the cone were sealed with light mineral oil.

**Figures**

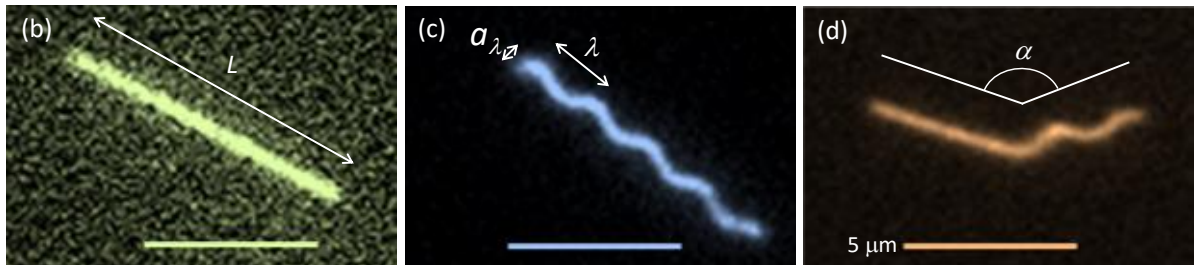

**Fig. 1. Straight, helical and rod-helix diblock flagellar filaments.** a) Fluorescence image of rigid flagellar filament with the contour length $L$. **b)** Helical flagellar filament with a wavelength $\lambda$=2.4 µm and a helical diameter $a_\lambda$=0.75 µm. c) Shape diblock copolymers composed of an adjoining rigid segment and a helical segment.

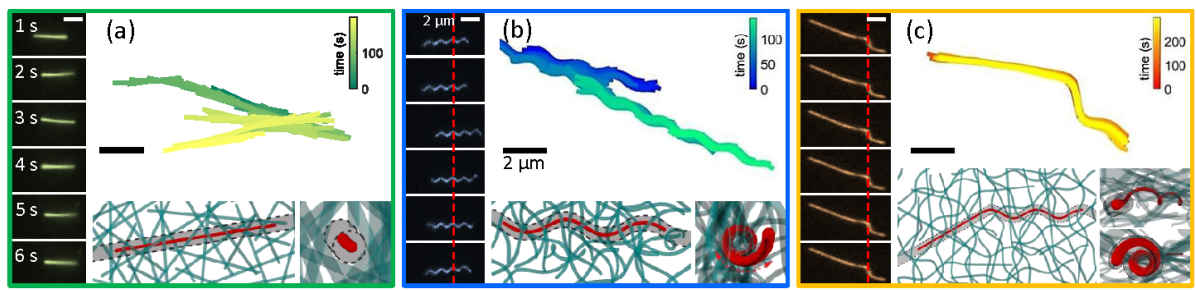

**Fig. 2. Filament shape determines dynamics in semi-dilute suspensions A)** Straight filaments preferentially diffuse along their long axis and slowly rotate, escaping their cage on a long time scale. **B)** Helical filaments exhibit corkscrewing dynamics where a translation of the filament's center of mass is tightly coupled to its rotation. For long times the helical filament escapes its cage by corkscrewing reptation-like dynamics. **C)** Helix-rod diblocks are permanently caged. In all panels a time sequence of fluorescently labelled filaments is shown on the left. The right panels correspond to tracked filament positions, with colors coding for time. Schematics of different filament shapes in the semi-dilute regime are shown on the bottom. Background suspensions concentration is 5 mg/mL.

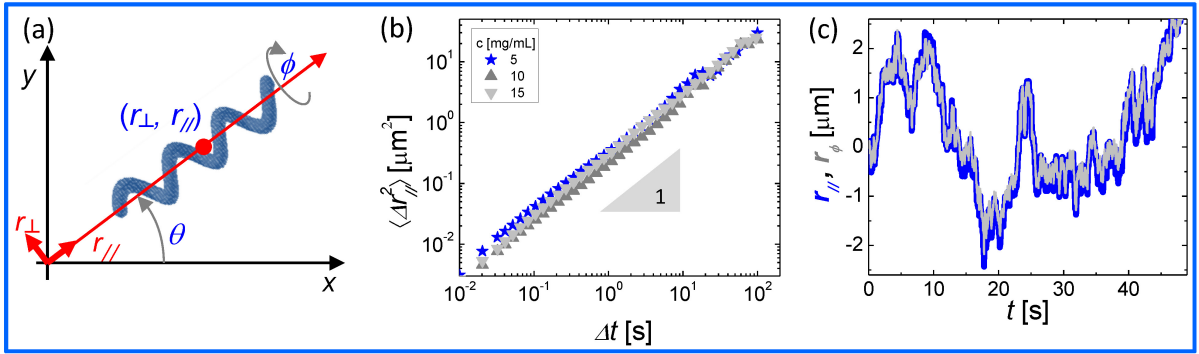

**Fig. 3. Corkscrewing dynamics of helical filaments. A)** Coordinates used to describe the motion of helical filaments. $\theta$ is the angle of between helix's long axis and the *x*-axis. $r_\parallel, r_\perp$ are displacements of the filament center of mass along and perpendicular to its long axis. $\phi$ is the phase angle of the filament's center of mass. **B)** The plot of MSD of helical filament along its long axis as a function of time. **C)** $r_\parallel$ and $r_\phi = \lambda\phi/2\pi$ quantitatively track each other as a function of time, demonstrating that long-term motion is dominated by corkscrewing dynamics.

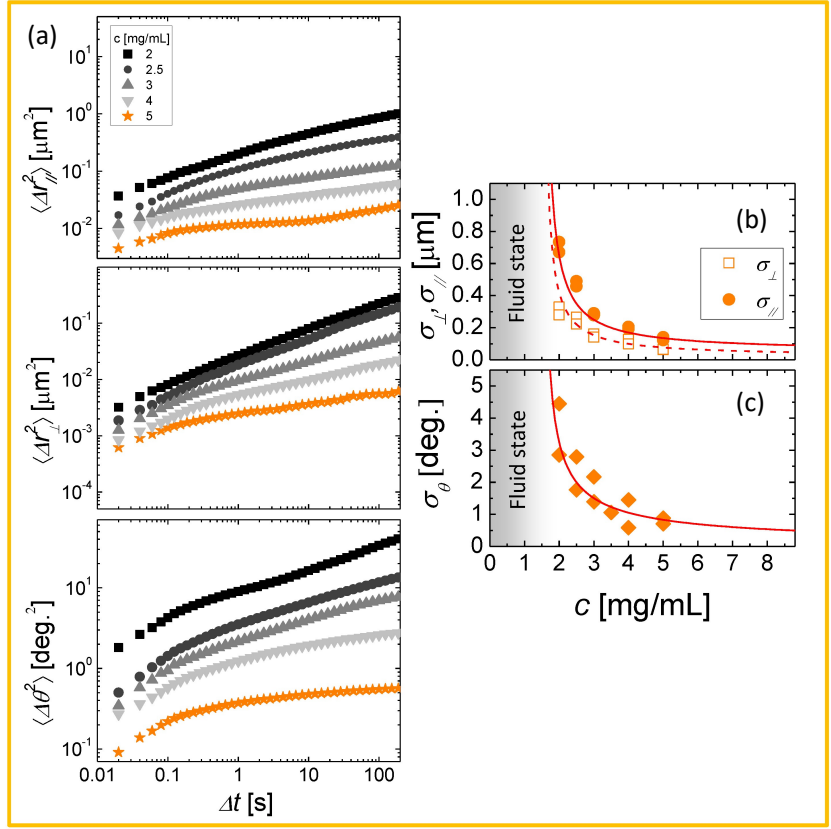

**Fig. 4. Dynamics of caged rod-helix diblock copolymers: A)** Mean Square Displacement (MSD) along ($r_{\parallel}$) and perpendicular ($r_{\perp}$) to the long axis as well as mean angular displacement as a function of the rod-helix concentration. **B-C)** Effective size of the filament cages extracted from the mean square spatial or angular displacement curves in panel a. $\sigma_{\parallel}$ $\sigma_{\perp}$, $\sigma_{\theta}$ are estimated from the average displacement at $\Delta t$=63 s. Red lines are a guide to the eye.

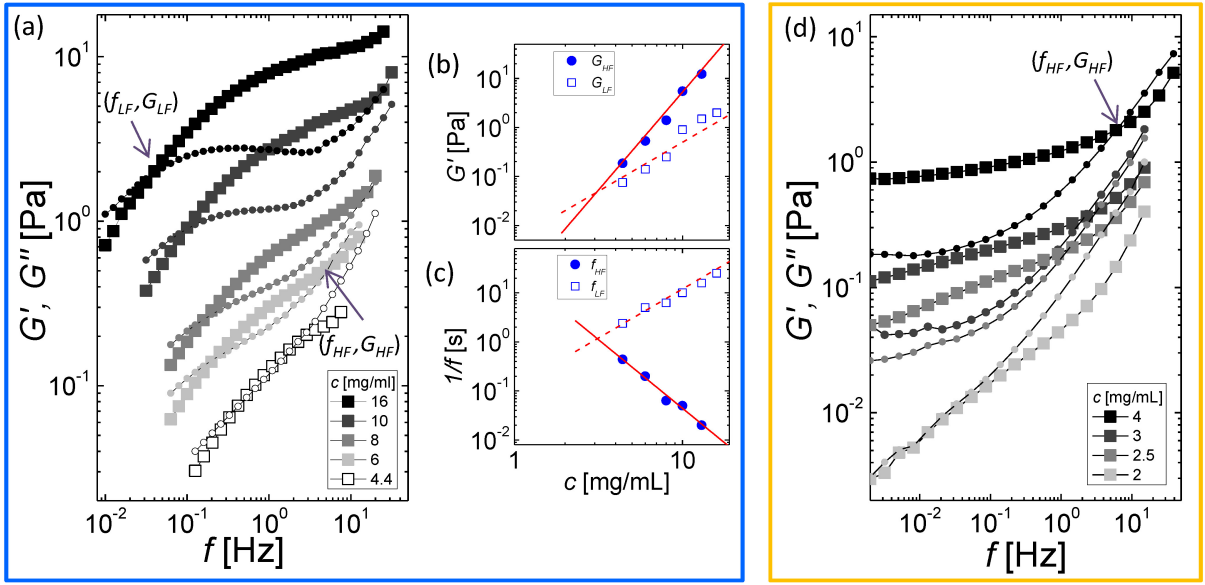

**Fig. 5. Rheology of helical and rod-helix diblock filament suspensions. A)** Storage (squares) and loss (circles) moduli as a function of frequency, for different helix concentrations. ($G_{LF}$ and $G_{HF}$ indicated low and high-frequency cross-over storage modulus. Cross-over characteristic times are given by $1/f_{LF}$ and $1/f_{HF}$. **B)** Elastic modulus at lower and higher cross-over frequency as a function of filament concentration. **C)** Cross over characteristic time $1/f_{HF}$ as a function of filament concentration. Red lines are guides to the eye. **D)** Storage (squares) and loss moduli (circles) of shape diblock suspensions for different filament concentrations.

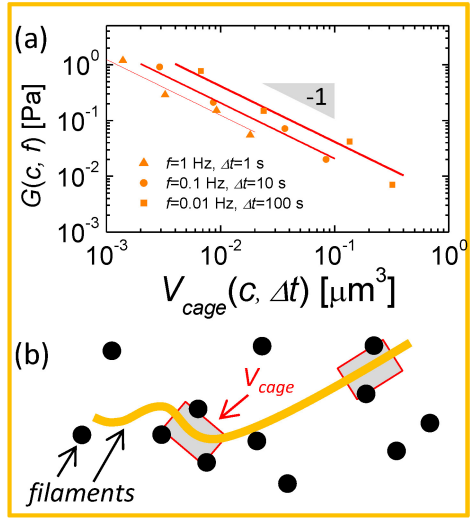

**Fig. 6. Comparison between rheology and the microscopic dynamics of diblock filaments. A)** $G_\infty$ as a function $V_{cage}$. ($V_{cage} = \pi(\sigma\perp/2)^2 \sigma_{//}$ ). **B)** Sketch of the dispersion. The dispersions behave as entropic glass. The crossover between the filaments is responsible for the dynamical arrest and the filament backbone ensure stress propagation in the dispersion.

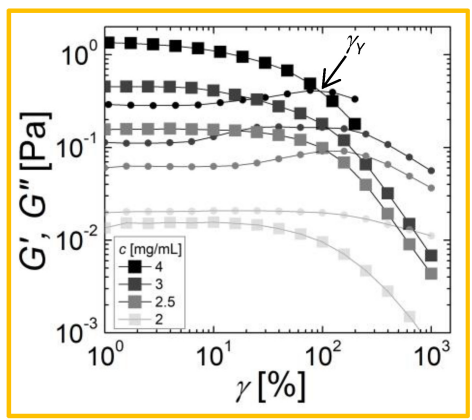

**Fig. 7. Strain amplitude sweeps of rod-helix suspensions.** Storage and loss modulus as a function of strain measured for a fixed frequency $\omega = 0.1\ Hz$.